\documentclass[conference]{IEEEtran}
\IEEEoverridecommandlockouts
\usepackage{cite}
\usepackage{amsmath,amssymb,amsfonts}
\usepackage{algorithmic}
\usepackage{graphicx}
\usepackage{textcomp}
\usepackage{xcolor}
\usepackage{multirow}
\usepackage{makecell}
\usepackage{url}
\usepackage{marvosym}

\def\BibTeX{{\rm B\kern-.05em{\sc i\kern-.025em b}\kern-.08em
    T\kern-.1667em\lower.7ex\hbox{E}\kern-.125emX}}
\begin{document}

\title{Enhanced Template-based Intra Mode Derivation with Adaptive Block Vector Replacement\\
\thanks{$^*$ denotes equal contribution. This work was supported in part by the National Science Foundation of China under grant no. 62025101 and grant no. 62501022, in part by the Science and Technology Plan Project in Basic Public Welfare class of Shaoxing city (No.2023A11006). (Corresponding to: Jiaqi Zhang, Jiaye Fu.)}
\thanks{}
}

\author{
\IEEEauthorblockN{Jiaqi Zhang$^{1,2}$$^\ast$, Jiaye Fu$^{3}$$^\ast$, Chuanmin Jia$^{1}$, Siwei Ma$^{1,2,3}$ \\ Karam Naser$^{4}$, Thierry Dumas$^{4}$, Saurabh Puri$^{4}$ and Miloš Radosavljevi\'{c}$^{4}$}
\IEEEauthorblockA{$^{1}$State Key Laboratory of Multimedia Information Processing, School of Computer Science, Peking University, China}
\IEEEauthorblockA{$^{2}$Information Technology R\&D Innovation Center, Peking University, Shaoxing, China}
\IEEEauthorblockA{$^{3}$School of Electronic and Computer Engineering, Peking University, China}
\IEEEauthorblockA{$^{4}$InterDigital R\&D, Cesson-Sévigné, France}
}

\maketitle

\begin{abstract}
Intra prediction is a crucial component in traditional video coding frameworks, aiming to eliminate spatial redundancy within frames. In recent years, an increasing number of decoder-side adaptive mode derivation methods have been adopted into Enhanced Compression Model (ECM). 
However, these methods predominantly rely on adjacent spatial information for intra mode decision-making, overlooking potential similarity patterns in non-adjacent spatial regions, thereby limiting intra prediction efficiency.
To address this limitation, this paper proposes a template-based intra mode derivation approach enhanced by block vector-based prediction. The adaptive block vector replacement strategy effectively expands the reference scope of the existing template-based intra mode derivation mode to non-adjacent spatial information, thereby enhancing prediction efficiency. Extensive experiments demonstrate that our strategy achieves 0.082\% Bjøntegaard delta rate (BD-rate) savings for Y components under the All Intra (AI) configuration compared to ECM-16.1 while maintaining identical encoding/decoding complexity, and delivers an additional 0.25\% BD-rate savings for Y components on screen content sequences.
\end{abstract}

\vspace{-0.2cm}
\section{Introduction}
\vspace{-0.1cm}
Versatile Video Coding~(VVC)~\cite{VVCoverview} represents the most recent video coding standard developed by the Joint Video Experts Team~(JVET), comprising ITU-T Video Coding Experts Group~(VCEG) and ISO/IEC Moving Picture Experts Group~(MPEG). 
VVC provides a significant enhancement over High Efficiency Video Coding~(HEVC)~\cite{HEVCoverview}, achieving greater than 50\% reduction in bit rate while preserving comparable subjective video quality. To pursue continued compression efficiency improvements, JVET established the Enhanced Compression Model~(ECM). The initial version, ECM-1.0, attained a 12\% Bjøntegaard delta rate (BD-rate)~\cite{BD-rate} saving under the Random Access~(RA) configuration. The latest iteration, ECM-16.1, achieves a substantial 27.06\% BD-rate improvement compared to the VVC reference implementation, VTM-11.0~\cite{AL0006}.

Intra prediction is a crucial component of conventional video coding frameworks, aimed at reducing spatial redundancy within frames. Intra prediction in ECM builds upon VVC foundations, including the 65 intra angular modes, Wide Angular Intra Prediction~(WAIP)~\cite{WAIP}, Intra Sub-Partition~(ISP)~\cite{ISP}, and Matrix-based Intra Prediction~(MIP)~\cite{MIP}. ECM enhances these existing tools and brings in new intra coding methods. Notably, ECM expands the Most Probable Modes~(MPMs) list from 6 to 23 angular modes~\cite{MPMS}. Additionally, ECM refines the Multiple Reference Line~(MRL)~\cite{MRL}, enabling more precise prediction based on multiple reference samples. 
It further introduces novel methodologies, such as the Extrapolation Filter-based Intra Prediction (EIP) mode~\cite{EIP}, designed to enhance prediction accuracy. The newly incorporated Intra Template Matching Prediction (IntraTMP)~\cite{IntraTMP} achieves more efficient prediction by reusing similar reconstructed regions within the frame. Furthermore, techniques leveraging spatial information from reconstructed regions for intra mode derivation are emerging as predominant intra prediction modes. Specifically, the Decoder-side Intra Mode Derivation~(DIMD)~\cite{DIMD} employs the Sobel operator to compute textural gradient information from neighboring spatial regions. Based on this analysis, it derives up to five intra prediction modes, whose prediction signals are subsequently fused via weighted intra fusion prediction. Additionally, the Occurrence-Based Intra-prediction Coding~(OBIC) mode~\cite{OBIC} statistically analyzes the frequency of prevalent intra prediction modes within the spatial domain and reutilizes up to five of these modes for intra fusion prediction. The Template-Based Intra Mode Derivation~(TIMD)~\cite{TIMD}, on the other hand, directly predicts the neighboring spatial template using angular modes. It then derives up to three optimal intra modes by minimizing the loss calculated against reconstructed pixel values, and finaly obtaining the prediction value through intra fusion process. Furthermore, the Spatial Geometric Partitioning Mode~(SGPM)~\cite{SGPM} directly partitions the template into corresponding segments. Subsequently, it derives a single optimal intra prediction mode for each segment. Finally, the prediction results from these two modes are fused via weighted averaging using predetermined partitioning weights. Collectively, these techniques significantly reduce coding overhead while maintaining prediction quality by exploiting information derived from already reconstructed spatial regions.

However, the recent development of intra prediction tools has encountered a bottleneck. This is primarily because existing intra tools predominantly utilize only adjacent spatial regions, which limits their prediction efficiency. Nevertheless, strategies involving non-adjacent spatial traversal, such as IntraTMP, introduce additional encoding and decoding complexity. To address these challenges, we design a flexible \textbf{Adaptive Block Vector Replacement} strategy and introduce an \textbf{Enhanced Template-based Intra Mode Derivation (E-TIMD)}. By traversing the block vectors used in both adjacent and non-adjacent intra spatial regions, the adaptive block vector replacement strategy first constructs a comprehensive list of intra block vector information. Subsequently, it performs adaptive mode replacement based on the evaluation metric of the E-TIMD. Concurrently, we extended the application of this list to DIMD, OBIC, and SGPM. This integration comprehensively enhances the reference scope of spatial information for these intra tools that utilize intra fusion prediction, thereby improving their prediction efficiency. Implemented on top of ECM-16.1~\cite{ECM}, the proposed method achieves 0.082\% BD-rate gain on the Y-component under All Intra configuration, with minor complexity changes. \textbf{The proposed E-TIMD with adaptive block vector replacement strategy is currently adopted in ECM.}

\vspace{-0.2cm}
\section{Related Work}
\vspace{-0.1cm}
This section provides an overview of the existing intra coding tools in ECM that are related to the proposed method. Specifically, TIMD-related tools are introduced in Section~\ref{sec::related:subsec:TIMD}. IntraTMP is discussed in Section~\ref{sec::related:subsec:IntraTMP}. 

\begin{figure}[t]
	\centering
	\includegraphics[width=0.6\columnwidth]{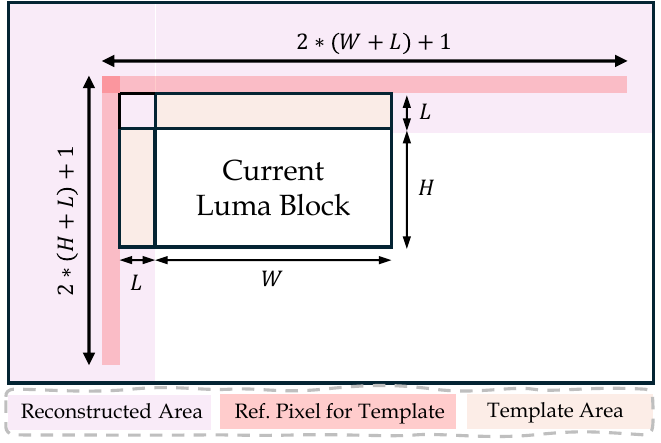}
        \vspace{-0.2cm}
	\caption{Illustration of TIMD.}
	\label{fig::timd}
        \vspace{-0.3cm}
\end{figure}
\begin{figure}[t]
	\centering
	\includegraphics[width=0.6\columnwidth]{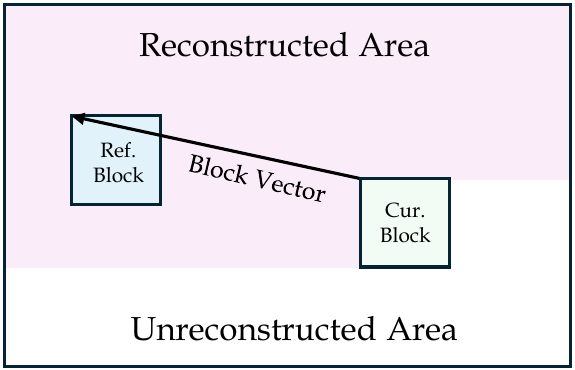}
        \vspace{-0.2cm}
	\caption{Illustration of IntraTMP.}
	\label{fig::intraTMP}
        \vspace{-0.4cm}
\end{figure}

\vspace{-0.2cm}
\subsection{Template-based Intra Mode Derivation}\label{sec::related:subsec:TIMD}

In ECM, when a coding block selects TIMD as its prediction method, the template area is first predicted using intra angular modes from the MPMs list, as illustrated in Figure~\ref{fig::timd}. Subsequently, template loss analysis calculates the Sum of Absolute Transform Differences (SATD) or Sum of Absolute Differences (SAD) between predicted and reconstructed pixel values within the template region. The intra angular mode with the lowest template loss is then selected for prediction. If the loss associated with the second-best intra angular mode is less than twice that of the best mode, both modes are utilized for TIMD prediction; otherwise, only the mode with the smallest loss is chosen. Furthermore, template loss analysis is performed to determine whether Planar or DC modes should be included as extra intra modes  for TIMD. The weight $w_{\sigma}$ assigned to the intra mode $\sigma$ involved in TIMD is inversely proportional to their corresponding template losses, which can be calculated by,
\begin{equation}
\footnotesize{
	\label{eqn:weightTimd}
	w _\sigma = \frac{\sum_{t \in \mathcal{T}}L_t}{\left( \mathcal{N}_{\mathcal{S}} - 1 \right) \times \sum_{s \in S}L_s} ,
}
\end{equation} 
where $S$ is the set of selected intra modes, $\mathcal{T}$ is the set $\mathcal{S}$ without intra mode $\sigma$, $L_t$ and $L_s$ represent the template loss of corresponding intra angular mode, and $\mathcal{N}_S$ is the number of intra angular mode in set $S$.
The final prediction for the luma block is obtained by the weighted fusion of predicted luma blocks using the selected intra modes, which is calculated by,
\begin{equation}
\footnotesize{
	\label{eqn:fusEq}
	p(x,y)= \sum_{m \in \mathbb{\mathcal{M}}}p_{m}(x,y)\times w_{m},
}
\end{equation}
where $p(x,y)$ is the ultimate prediction result located at $(x,y)$ of the current block, 
$\mathbb{\mathcal{M}}$ is the set of selected intra modes, 
$p_{m}(x,y)$ is the prediction result of intra angular mode $m$, and $w_{m}$ is the calculated weight for intra angular mode $m$.

\vspace{-0.2cm}
\subsection{Intra Template Matching Prediction}\label{sec::related:subsec:IntraTMP}
IntraTMP innovatively adapts and extends the well-established principle of template matching, which is commonly used in inter prediction for eliminating temporal redundancy, to the domain of intra prediction. Unlike traditional intra prediction modes that utilize adjacent pixels in predefined directions, IntraTMP searches within a broader spatial context. The core process begins by defining a specific search window within the already reconstructed neighboring region of the current block. A distinct feature of IntraTMP is its use of a $\Gamma$-shaped template, which encapsulates both the top and left neighboring pixels of the current luma block.
This $\Gamma$-shaped template then scans across the designated search area. The objective is to identify the reference block whose $\Gamma$-shaped template exhibits the highest similarity to the template of the current block. Typically, a cost function like SAD or SATD is employed to quantify this similarity. A relative displacement coordinates, which are often referred to as a Block Vector~(BV), indicate location of an optimal matching reference block.

Subsequently, during the prediction phase for the current luma block, the stored BV is retrieved. The pixel values within the current block are then predicted by replicating the corresponding pixel values from the identified reference block in the reconstructed region. This mechanism effectively leverages complex, non-directional spatial correlations that may occur across distant regions within the frame, as illustrated in Figure~\ref{fig::intraTMP}.

\vspace{-0.2cm}
\section{Methodology}
\vspace{-0.1cm}
In this section, we provide an in-depth elaboration of the proposed E-TIMD and the Adaptive Block Vector Replacement Strategy. Specifically, Section~\ref{sec::method:subsec:consBVL} introduces the construction methodology of the Block Vector (BV) list. Subsequently, Section~\ref{sec::method:subsec:ETIMD} delineates the operational mechanism of the Adaptive Block Vector Replacement Strategy within the E-TIMD framework. Finally, Section~\ref{sec::method:subsec:furtheropt} then introduces supplementary optimizations related to the E-TIMD.

\vspace{-0.1cm}
\subsection{The Construction of BV List}\label{sec::method:subsec:consBVL}
\begin{figure}[t]
	\centering
	\includegraphics[width=0.55\columnwidth]{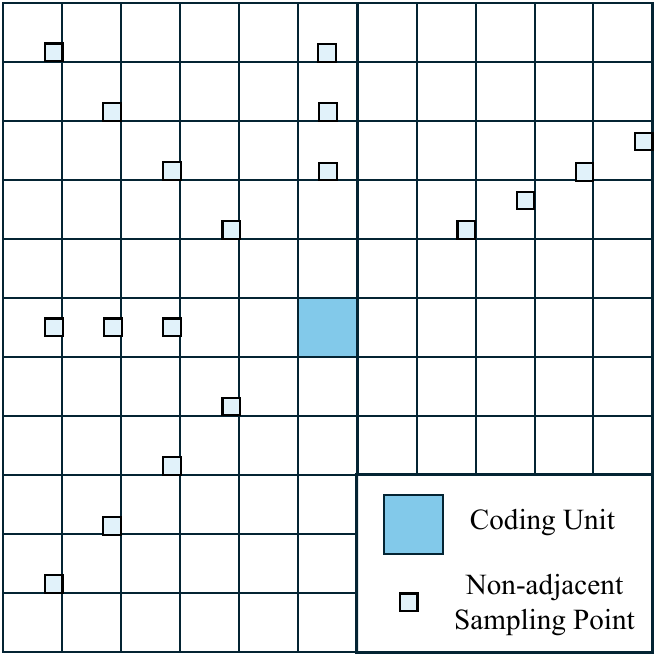}
        \vspace{-0.2cm}
	\caption{Spatial Sampling of Block Vector.}
	\label{fig::sampling}
        \vspace{-0.3cm}
\end{figure}
\begin{figure}[t]
	\centering
	\includegraphics[width=0.5\columnwidth]{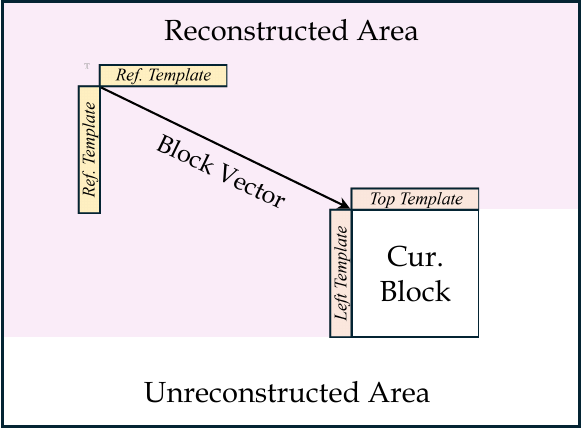}
        \vspace{-0.2cm}
	\caption{The Template Prediction through Block Vector.}
	\label{fig::template-copy}
        \vspace{-0.4cm}
\end{figure}
Since exhaustively traversing the reconstructed area to construct a Block Vector (BV) list would impose prohibitive computational complexity on both encoder and decoder, the Adaptive Block Vector Replacement Strategy traverses predefined sampling points across adjacent and non-adjacent spatial domains, with the non-adjacent spatial sampling shown in Figure~\ref{fig::sampling}.  
When a sampled point corresponds to a reconstructed coding unit using IntraTMP or Intra Block Copy as its prediction method, the associated BVs undergo mandatory integer-pixel normalization according to their coding type before integration into the candidate list.
Concurrently, for coding units utilizing DIMD, OBIC, SGPM, or E-TIMD, any block vector mode participating in their intra fusion prediction process contributes its BV to the candidate list.

Upon completing this primary BV acquisition phase, a secondary refinement stage inspired by Auto Relocated-BV~(AR-BV) methodology~\cite{AR-BV-1} is activated: Each BV within the preliminary list propagates additional candidates when its referenced block itself employs either block vector prediction or intra fusion prediction. Notably, the corresponding AR-BV is also added into the list, which is calculated through:
\begin{equation}
    AR\_BV=BV_{pri} + BV_{sec},
\end{equation}
where $BV_{pri}$ is the initial block vector from the primary candidate list, and $BV_{sec}$ is the block vector extracted from the prediction unit referenced by $BV_{pri}$. This secondary mechanism effectively exploits spatial dependencies across prediction hierarchies, thereby culminating in a comprehensive, multi-sourced BV list optimized for both prediction accuracy and computational efficiency.

\vspace{-0.2cm}
\subsection{Enhanced Template-based Intra Mode Derivation}\label{sec::method:subsec:ETIMD}
Following acquisition of the complete BV list, selecting candidate modes from both intra angular modes and block vector modes in E-TIMD becomes a key challenge.
Conventional approaches that first determine original intra angular modes in TIMD and then perform adaptive block vector mode replacement prove suboptimal for prediction efficiency.
This limitation stems from the strict dependence of the number of intra angular modes participating in TIMD on template cost analysis. Consequently, we fundamentally redesign TIMD's mode selection framework to seamlessly integrate the Adaptive Block Vector Replacement Strategy.

E-TIMD initiates the process by copying the template associated with the BV to the current block according to the methodology illustrated in Figure~\ref{fig::template-copy}, followed by computing a matching cost (e.g., SAD or SATD). Subsequently, after obtaining the costs for conventional intra angular modes, all costs are sorted in ascending order. The mode yielding the minimum cost is selected as the primary prediction mode for E-TIMD.
A secondary mode is incorporated into E-TIMD if the second-lowest cost satisfies the following condition:
\begin{equation}
\label{eq:loss_condition}
\mathcal{L}_{sec} < 1.5 \times \mathcal{L}_{fir}
\end{equation}
where $\mathcal{L}_{fir}$ denotes the cost of the primary E-TIMD mode and $\mathcal{L}_{sec}$ represents the cost of the second-ranked candidate.
When two modes are selected for E-TIMD, the third mode is adaptively chosen from a candidate pool with the lowest cost, with candidate pool comprising the Planar mode, DC mode, and BV modes.
Thus, the E-TIMD mode selection framework systematically incorporates both intra angular modes and BV modes to maximize prediction effectiveness. Finally, to ensure design homogeneity within the ECM and facilitate standardization, all E-TIMD parameters, including mode weights, location depth, and prediction methodologies, strictly adhere to the conventional TIMD implementation.

\begin{table*}[!t]
\centering
\caption{Coding performance of E-TIMD \\ (BD-rates and relative runtimes for ECM-16.1 CTC).}
\label{tab::ai-result1}
\vspace{-0.2cm}
\begin{tabular*}{0.85\textwidth}{@{\extracolsep{\fill}}cc|ccccccc}
\hline
\hline
\multirow{2}{*}{Class} &\multirow{2}{*}{Sequences} &\multicolumn{7}{c}{\textbf{All Intra Main 10}}\\
                  & &Y &U & V & $\Delta$EncT & $\Delta$DecT &EncVmPeak &DecVmPeak \\
\hline
\multirow{3}{*}{  \makecell[c]{A1\\3840$\times$2160}  } 
& Tango           & -0.04\% &  0.39\% &  0.21\% & 100.56\% & 100.06\% & 100.67\% & 100.89\%\\
& FoodMarket      & -0.01\% &  0.01\% & -0.17\% & 99.79\%  & 100.70\% & 100.73\% & 101.12\%\\
& CampFire        & -0.01\% & -0.10\% & -0.12\% & 98.71\%  & 102.20\% & 100.86\% & 101.51\%\\
\hline
\multirow{3}{*}{ \makecell[c]{A2\\3840$\times$2160} } 
& CatRobot        & -0.07\% & -0.17\% & -0.16\% & 98.41\%  & 102.57\% & 100.84\% & 101.38\%\\
& DaylightRoad    & -0.19\% & -0.34\% & -0.16\% & 99.54\%  & 100.26\% & 100.98\% & 101.75\%\\
& ParkRunning     &  0.00\% & -0.03\% & -0.04\% & 99.34\%  & 100.15\% & 100.74\% & 101.13\%\\
\hline
\multirow{5}{*}{ \makecell{B\\1920$\times$1080} }  
& MarketPlace     & -0.01\% & -0.10\% & -0.18\% & 100.30\% & 101.95\% & 100.44\% & 101.75\%\\
& RitualDance     & -0.05\% &  0.10\% &  0.08\% &  99.11\% & 100.97\% & 100.37\% & 100.76\%\\
& Cactus          & -0.07\% & -0.04\% &  0.06\% &  99.79\% & 101.87\% & 100.49\% & 101.05\%\\
& BasketballDrive & -0.09\% &  0.05\% & -0.19\% & 100.38\% & 100.39\% & 100.51\% & 101.63\%\\
& BQTerrace       & -0.19\% & -0.18\% & -0.19\% & 100.01\% & 100.49\% & 100.57\% & 101.34\%\\
\hline
\multirow{4}{*}{ \makecell{C\\832$\times$480} }  
& BasketballDrill & -0.14\% &  0.13\% &  0.02\% & 101.15\% & 100.93\% & 100.54\% & 100.57\%\\
& BQMall          & -0.04\% &  0.15\% & -0.05\% &  98.74\% & 101.23\% & 107.88\% & 100.48\%\\
& PartyScene      & -0.20\% &  0.03\% & -0.08\% & 100.34\% & 101.45\% & 100.62\% & 100.61\%\\
& RaceHorses      &  0.00\% & -0.35\% & -0.06\% &  99.43\% & 100.96\% & 100.57\% & 100.43\%\\
\hline
\multirow{3}{*}{  \makecell{E\\1280$\times$720}  }  
& FourPeople      & -0.15\% &  0.03\% &  0.12\% &  99.08\% & 100.47\% & 100.44\% & 100.31\%\\
& Johnny          & -0.06\% & -0.02\% & -0.48\% &  99.92\% & 100.39\% & 100.38\% & 100.18\%\\
& KristenAndSara  & -0.14\% & -0.09\% &  0.11\% &  99.43\% & 100.41\% & 100.41\% & 100.29\%\\
\hline
\hline
& \textbf{Overall}&\textbf{ -0.081\% }&\textbf{ 0.009\% }&\textbf{ -0.070\% }&\textbf{ 99.67\% }&\textbf{ 101.00\%}&\textbf{ 100.67\%}&\textbf{ 100.89\% }\\
\hline
\hline
\end{tabular*}
\end{table*}
\begin{table*}[!t]
\centering
\caption{Coding performance of E-TIMD with optimization of transform coefficient selection \\ (BD-rates and relative runtimes for ECM-16.1 CTC).}
\label{tab::ai-result2}
\vspace{-0.2cm}
\begin{tabular*}{0.85\textwidth}{@{\extracolsep{\fill}}cc|ccccccc}
\hline
\hline
\multirow{2}{*}{Class} &\multirow{2}{*}{Sequences} &\multicolumn{7}{c}{\textbf{All Intra Main 10}}\\
                     &               &Y &U & V &$\Delta$EncT & $\Delta$DecT &EncVmPeak &DecVmPeak \\
\hline
\multirow{3}{*}{  \makecell[c]{A1\\3840$\times$2160}  } 
& Tango           & -0.05\% & 0.58\% & 0.09\% & 100.40\% & 101.99\% & 100.81\% & 100.97\% \\
& FoodMarket      & -0.02\% & -0.04\% & 0.01\% & 100.60\% & 99.37\% & 100.86\% & 101.34\% \\
& Campfire        & -0.01\% & -0.06\% & -0.12\% & 101.46\% & 100.76\% & 100.90\% & 101.72\% \\
\hline
\multirow{3}{*}{ \makecell[c]{A2\\3840$\times$2160} } 
& CarRobot        & -0.07\% & -0.03\% & -0.17\% & 101.00\% & 101.32\% & 100.95\% & 101.44\% \\
& DayLightRoad    & -0.21\% & -0.08\% & -0.11\% & 99.87\% & 100.60\% & 101.11\% & 102.15\% \\
& ParkRunning     & 0.00\% & -0.04\% & -0.04\% & 100.40\% & 100.61\% & 100.87\% & 101.32\% \\
\hline
\multirow{5}{*}{ \makecell{B\\1920$\times$1080} }  
& MarketPlace     & -0.02\% & -0.20\% & 0.00\% & 100.66\% & 101.70\% & 100.47\% & 100.52\% \\
& RitualDance     & -0.03\% & -0.05\% & -0.01\% & 100.61\% & 100.57\% & 100.50\% & 101.03\% \\
& Cactus          & -0.07\% & -0.01\% & 0.03\% & 100.35\% & 100.01\% & 100.56\% & 101.11\% \\
& BasketballDrive & -0.12\% & -0.15\% & -0.17\% & 100.77\% & 100.08\% & 100.59\% & 101.87\% \\
& BQTerrace       & -0.20\% & -0.03\% & -0.18\% & 100.82\% & 100.68\% & 100.69\% & 101.58\% \\
\hline
\multirow{4}{*}{ \makecell{C\\832$\times$480} }  
& BasketballDrill & -0.07\% & 0.12\% & 0.04\% & 103.32\% & 100.63\% & 100.70\% & 100.36\% \\
& BQMall          & -0.04\% & 0.16\% & 0.20\% & 100.43\% & 101.09\% & 101.83\% & 100.56\% \\
& PartyScene      & -0.20\% & -0.04\% & -0.15\% & 100.92\% & 100.90\% & 100.74\% & 100.86\% \\
& RaceHorses      & 0.02\% & 0.11\% & -0.14\% & 99.74\% & 101.24\% & 100.63\% & 100.36\% \\
\hline
\multirow{3}{*}{  \makecell{E\\1280$\times$720}  }  
& FourPeople      & -0.16\% & -0.02\% & -0.18\% & 100.31\% & 100.47\% & 100.50\% & 100.33\% \\
& Johnny          & -0.06\% & 0.05\% & -0.40\% & 100.97\% & 100.81\% & 100.51\% & 100.28\% \\
& KristenAndSara  & -0.15\% & -0.01\% & 0.24\% & 101.40\% & 100.55\% & 100.49\% & 100.11\% \\
\hline
\hline
& \textbf{Overall}&\textbf{ -0.082\% }&\textbf{ 0.013\% }&\textbf{ -0.059\% }&\textbf{ 100.78\% }&\textbf{ 100.74\% }&\textbf{ 100.76\% }&\textbf{ 100.99\% }\\
\hline
\hline
\multirow{4}{*}{  \makecell{F\\1280$\times$720\\1920$\times$1080} }  
& BasketballDrillText   & -0.10\% & -0.18\% & -0.05\% & 101.83\% & 101.23\% & 107.84\% & 100.48\% \\
& ArenaOfValor          & -0.10\% & -0.11\% & -0.09\% & 100.68\% & 100.81\% & 100.50\% & 101.40\% \\
& SlideEditing          & -0.34\% & -0.15\% & -0.37\% & 100.00\% & 101.10\% & 100.56\% & 100.81\% \\
& SlideShow             & -0.15\% & -0.24\% &  0.12\% & 100.48\% & 101.13\% & 100.43\% & 100.10\% \\
\hline
\multirow{4}{*}{ \makecell{TGM\\1920$\times$1080} }  
& FlyingGraphic    & -0.40\% & -0.24\% & -0.24\% & 103.53\% & 100.77\% & 101.07\% & 100.90\% \\
& Desktop          & -0.36\% & -0.27\% & -0.30\% & 100.24\% & 101.03\% & 100.48\% & 100.41\% \\
& Console          & -0.23\% & -0.20\% & -0.18\% & 100.04\% & 100.82\% & 100.51\% & 101.05\% \\
& ChineseEditing   & -0.35\% & -0.34\% & -0.34\% & 100.15\% & 100.96\% & 100.56\% & 101.16\% \\
\hline
\hline
& \textbf{Overall (SCC)} &\textbf{-0.25\%}&\textbf{-0.21\%}&\textbf{-0.18\%}&\textbf{-100.9\%}&\textbf{101.0\%}&\textbf{101.5\%}&\textbf{100.8\%}\\
\hline
\hline
\end{tabular*}
\end{table*}
\vspace{-0.2cm}
\subsection{Further Optimization of E-TIMD}\label{sec::method:subsec:furtheropt}
When a BV mode serves as the primary prediction mode in E-TIMD, transform coefficient selection becomes challenging.
Conventionally, coding units using block vector prediction uniformly adopt transform coefficient selection identical to that of Planar mode.
However, this coefficient assignment proves to be suboptimal. Consequently, E-TIMD introduces a dedicated transform coefficient selection strategy specifically designed for cases where BV-based modes function as the primary prediction mechanism.

This strategy optimizes transform coefficient selection by \textit{correlating intra-block textural gradients with intra angular modes.} Specifically, each coding unit constructs a Histogram of Gradients (HoG) through sliding-window edge detection. For each window, the gradient orientation $\theta$ is computed by:
\begin{equation}
    \theta =\arctan \left ( \frac{G_{ver}}{G_{hor}}  \right ),
\end{equation}
where $G_{ver}$ and $G_{hor}$ are the vertical gradient and horizontal gradient respectively. 
These orientations are quantized to align with ECM's 65 intra angular modes, with corresponding frequency counts accumulated in the HoG. The angular mode exhibiting peak HoG frequency is identified as the optimal textural direction. Within the E-TIMD's framework, where up to two prediction modes participate in transform selection, any BV mode is substituted with this HoG-determined optimal angular mode. This ensures transform coefficient selection maintain compatibility with the block's intrinsic directional characteristics, thereby enhancing coding efficiency.

\vspace{-0.2cm}
\section{Experiments}
\vspace{-0.1cm}
To evaluate the proposed E-TIMD method's performance, we integrated it into the ECM-16.1 reference software.
Simulations followed JVET Common Test Conditions (CTC) and evaluation procedures for enhanced compression tools~\cite{CTC}. Results were obtained under the All-Intra configuration using Quantization Parameters (QPs) of 22, 27, 32, and 37. Coding performance was evaluated using BD-rate, where negative values indicate compression gains. Encoding and decoding complexity, denoted by $\bigtriangleup EncT$ and $\bigtriangleup DecT$ respectively, are defined as:

\begin{equation}
\footnotesize{
\label{eqn:encT}
\bigtriangleup EncT=\frac{TEnc_{Proposed}}{TEnc_{Anchor}}\times 100
}
\end{equation}
\begin{equation}
\footnotesize{
\label{eqn:decT}
\bigtriangleup DecT=\frac{TDec_{Proposed}}{TDec_{Anchor}}\times 100
}
\end{equation}
where $TEnc_{Proposed}$ and $TDec_{Proposed}$ are the encoding and decoding time of the proposed method, and $TEnc_{Anchor}$ and $TDec_{Anchor}$ represent the encoding and decoding time of the anchor, i.e., ECM-16.1. Moreover, in the current test sequences, class A1, A2, B, C, and E are natural content sequences with different resolutions (720p to 4K), and the average BD-rate reduction is obtained through natural content sequences. Class F and TGM are screen content sequences characterized primarily by relatively repetitive, high-frequency textures.


As presented in Table~\ref{tab::ai-result1}, the proposed E-TIMD achieves BD-rate reductions of 0.081\%, -0.009\%, and 0.070\% for the Y, Cb, and Cr components, respectively, compared to the ECM-16.1. In Table~\ref{tab::ai-result2}, the Y-component BD-rate savings can be increased to 0.082\% when the optimization of transform coefficient selection is switched on. In addition, the complexity on the codec side of the optimized E-TIMD is less than 101\%, indicating that our technique can bring almost no additional complexity on both encoder side and decoder side. However, since extra memory is required to keep the BV list of each CU, the virtual memory peaks are slightly increased at both sides.

Moreover, from the table, it can be seen that the introduction of E-TIMD brings approximately 0.25\% luma gain for screen content sequences. This demonstrates that our method effectively enhances the existing ECM's compression capability for screen content. The above results provide innovative insights for the development of next-generation coding standards.


\vspace{-0.2cm}
\section{Conclusion}
\vspace{-0.1cm}
This paper introduces Enhanced Template-based Intra Mode Derivation (E-TIMD) with a plug-and-play Adaptive Block Vector Replacement strategy for luma intra prediction.
The adaptive block vector replacement strategy efficiently captures non-adjacent spatial information through comprehensive block vector search.
Combined with precisely engineered enhanced template-based intra mode derivation, this work proposes the first unified intra prediction framework that jointly leverages both adjacent and non-adjacent spatial information.
Furthermore, for E-TIMD, a transform coefficient selection algorithm is developed, which leverages reconstructed block texture gradients to optimize coefficient selection and enhance prediction efficiency.
Experimental results demonstrate that compared to ECM-16.1, the proposed E-TIMD achieves 0.082\% BD-rate savings while introducing no additional encoding/decoding latency. 
Moreover, the method incurs negligible virtual memory usage increase at both encoder and decoder. The comprehensive performance characterization presented in this work hold significant implications for next-generation video coding standardization.


\begin{thebibliography}{5}
\bibitem{VVCoverview}
B.~Bross, Y.-K. Wang, Y.~Ye, S.~Liu, J.~Chen, G.~J. Sullivan, and J.-R. Ohm, ``{Overview of the Versatile Video Coding (VVC) Standard and its Applications},'' \emph{IEEE Transactions on Circuits and Systems for Video Technology}, vol.~31, no.~10, pp. 3736--3764, 2021.

\bibitem{HEVCoverview}
G.~J. Sullivan, J.-R. Ohm, W.-J. Han, and T.~Wiegand, ``{Overview of the High Efficiency Video Coding (HEVC) Standard},'' \emph{IEEE Transactions on Circuits and Systems for Video Technology}, vol.~22, no.~12, pp. 1649--1668, 2012.

\bibitem{BD-rate}
G.~Bj{\o}ntegaard, ``Calculation of average {PSNR} differences between {RD}-curves ({VCEG-M33}),'' in \emph{VCEG Meeting (ITU-T SG16 Q. 6)}, 2001, pp. 2--4.

\bibitem{AL0006}
V.~Seregin, J.~Chen, R.~Chernyak, F.~Le~Leannec, and K.~Zhang, ``{JVET AHG report: ECM software development (AHG6)},'' \emph{document JVET-AL0006}, 2025.

\bibitem{WAIP}
L.~Zhao, X.~Zhao, S.~Liu, X.~Li, J.~Lainema, G.~Rath, F.~Urban, and F.~Racap{\'e}, ``{Wide Angular Intra Prediction for Versatile Video Coding},'' in \emph{Data Compression Conference (DCC)}.\hskip 1em plus 0.5em minus 0.4em\relax IEEE, 2019, pp. 53--62.

\bibitem{ISP}
S.~De-Lux{\'a}n-Hern{\'a}ndez, V.~George, J.~Ma, T.~Nguyen, H.~Schwarz, D.~Marpe, and T.~Wiegand, ``{An Intra Subpartition Coding Mode for VVC},'' in \emph{International Conference on Image Processing (ICIP)}.\hskip 1em plus 0.5em minus 0.4em\relax IEEE, 2019, pp. 1203--1207.

\bibitem{MIP}
M.~Sch{\"a}fer, B.~Stallenberger, J.~Pfaff, P.~Helle, H.~Schwarz, D.~Marpe, and T.~Wiegand, ``{An Affine-linear Intra Prediction with Complexity Constraints},'' in \emph{International Conference on Image Processing (ICIP)}.\hskip 1em plus 0.5em minus 0.4em\relax IEEE, 2019, pp. 1089--1093.

\bibitem{MPMS}
V.~Seregin, W.-J. Chien, M.~Karczewicz, and N.~Hu, ``{Block Shape Dependent Intra Mode Coding},'' \emph{document JVET-D0114}, 2016.

\bibitem{MRL}
K.~Cao, Y.-J. Chang, B.~Ray, V.~Seregin, M.~Karczewicz, and N.~Hu, ``{Non-EE2: Extended MRL Candidate List},'' \emph{document JVET-X0142}, 2021.

\bibitem{EIP}
L.~Xu, Y.~Yu, H.~Yu, and D.~Wang, ``{Non-EE2: An Extrapolation Filter-based Intra Prediction Mode},'' \emph{document JVET-AD0081}, 2023.

\bibitem{IntraTMP}
G.~Venugopal, K.~Müller, J.~Pfaff, H.~Schwarz, D.~Marpe, and T.~Wiegand, ``{Region-Based Template Matching Prediction for Intra Coding},'' \emph{IEEE Transactions on Image Processing}, vol.~32, pp. 779--790, 2023.

\bibitem{DIMD}
A.~Nasrallah, E.~Mora, T.~Guionnet, and M.~Raulet, ``{Decoder-Side Intra Mode Derivation Based on a Histogram of Gradients in Versatile Video Coding},'' in \emph{Data Compression Conference (DCC)}, 2019, pp. 597--597.

\bibitem{OBIC}
R.~G. Youvalari and M.~Abdoli, ``{AHG 12: Occurrence-Based Intra Coding (OBIC)},'' \emph{document JVET-AG0141}, 2024.

\bibitem{TIMD}
Y.~Wang, L.~Zhang, K.~Zhang, Z.~Deng, and N.~Zhang, ``{EE2-related: Template-based Intra Mode Derivation Using MPMs},'' \emph{document JVET-V0098}, 2021.

\bibitem{SGPM}
F.~Wang, Y.~Yu, H.~Yu, and D.~Wang, ``{Non-EE2: Spatial GPM},'' \emph{document JVET-Z0124}, 2022.

\bibitem{ECM}
``{ECM-16.1} software repository,'' \url{https://vcgit.hhi.fraunhofer.de/ecm/ECM/-/tree/ECM-16.1}.

\bibitem{AR-BV-1}
N.~Zhang, K.~Zhang, and L.~Zhang, ``{Non-EE2: Auto Relocated Block Vector Prediction},'' \emph{document JVET-AF0129}, 2023.

\bibitem{CTC}
M.~Karczewicz and Y.~Ye, ``{Common Test Conditions and Evaluation Procedures for Enhanced Compression Tool Testing},'' \emph{document JVET-AI2017}, 2024.

\end{thebibliography}

\end{document}